# Lattice dynamics and thermodynamics of bcc iron at pressure: first-principles linear response study

Xianwei Sha and R. E. Cohen

Carnegie Institution of Washington, 5251 Broad Branch Road, NW, Washington, DC 20015

**ABSTRACT**

We compute the lattice-dynamical and thermal equation of state properties of ferromagnetic bcc iron using the first principles linear response linear-muffin-tin-orbital method in the generalized-gradient approximation. The calculated phonon dispersion and phonon density of states, both at ambient and high pressures, show good agreement with inelastic neutron scattering data. We find the free energy as a function of volume and temperature, including both electronic excitations and phonon contributions, and we have derived various thermodynamic properties at high pressure and temperature. The thermal equation of state at ambient temperature agrees well with diamond-anvil-cell measurements. We have performed detailed investigations on the behavior of various thermal equation of state parameters, such as the bulk modulus K, the thermal expansivity α, the Anderson-Grüneisen parameter $\delta_T$, the Grüneisen ratio γ, and the heat capacity $C_V$ as function of temperature and pressure. A detailed comparison has been made with available experimental measurements, as well as results from similar theoretical studies on nonmagnetic bcc Tantalum.

# 1. Introduction

During the past decade, tremendous experimental[1-7] and theoretical[8-14] efforts have been devoted to investigate various properties of iron, especially for those at high pressure and temperature conditions. Body-center-cubic (bcc) is the ground state structure for iron at ambient conditions. It transforms to face-center-cubic (fcc) structure at elevated temperature (~1150K at ambient pressure), and to hexagonal-close-packed (hcp) structure at increased pressure (~11GPa at room temperature)[15, 16]. Iron also shows very interesting and complex magnetic behavior under different temperature and pressure conditions. It has become a fundamental problem in material science to understand the mechanism of these solid-state phase transitions, the magnetism, and the phase diagram of Fe. The study of iron is also of great geophysical interest, because the Earth's core consists mainly of this element.

Various lattice dynamical and thermodynamic properties as a function of temperature and pressure may provide important information to understand the phase transitions, phase diagram and dynamic response of materials. Several first-principle calculations[13, 14, 17] have been performed to study the lattice dynamical properties of bcc iron, generally giving good agreement with experiment[18]. Here we concentrate on the thermal equation of state of ferromagnetic bcc iron, using quasiharmonic lattice dynamics with an all-electron method that does not depend on pseudopotentials.

In section II we detailed our methods to perform first principles calculations, as well as the theoretical techniques to obtain thermal properties. We present our results and related discussions about lattice dynamics in section III, and about the thermal equation of state properties in section IV. We conclude with a summary in Section V.

## 2. Theoretical methods

For many metals and alloys, the Helmholtz free energy F of a solid has three major contributions[19]:

$$F(V,T)=E_{static}(V)+F_{el}(V,T)+F_{vib}(V,T) \qquad (1)$$

With V as the volume, and T as the temperature. $E_{static}(V)$ is the energy of a static lattice at absolute zero temperature, $F_{el}(v,T)$ is the electronic thermal free energy arising from electronic excitations, and $F_{vib}(V,T)$ is the vibrational energy contribution. We assume that the existence of lattice vibrations does not significantly affect the electronic contribution for the thermal properties, and all the terms here are calculated for the ideal lattice. $E_{static}(V)$ and $F_{el}(v,T)$ can be obtained from first-principles calculations directly. There are several ways to examine the lattice vibrational contribution, including the linear response (LR) lattice dynamics, particle-in-cell (PIC) model and molecular dynamics. First-principles LR calculations can give important lattice dynamics information, but it is computationally expensive. Additionally, it usually requires use of the quasi-harmonic approximation, and anharmonic effects are usually neglected. The PIC model is a mean field approximation to the thermal contribution[20, 21]. It includes on-site anharmonicity and requires less computer power when combined with first-principles total-energy calculations. However, it neglects the interatomic correlations and diffusion. The accuracy and reliability of molecular dynamics methods strongly depends on the quality and versatility of the interatomic potential. It is computational expensive to study the thermal equation of state properties rigorously from first-principles molecular dynamics calculations directly, and one obtains only the classical contributions, so that properties at room temperature and below are not reliable. In the present paper, we use first principles full potential Linear-Muffin-Tin-Orbital (LMTO) calculations to evaluate $E_{static}(V)$ and $F_{el}(v,T)$, and linear response LMTO calculations for the lattice vibrational properties.

Since bcc Fe is ferromagnetic, magnetic fluctuations at high temperatures contribute to the Helmholtz free energies. We are currently performing Monte Carlo simulations based on an effective Hamiltonian to examine the contribution from magnetic fluctuations in bcc Fe, and the results will be reported in the future.

A. Static lattice energy

We use multi-κ basis sets and two energy panels in the LMTO method[22, 23]. Space is divided into the non-overlapping Muffin-Tin (MT) spheres surrounding each individual atom and the remaining interstitial region. Non-overlapping muffin-tin spheres with radii of 2.18 bohr have been used in all the calculations. The self-consistent calculations are performed using 3κ-*spd*-LMTO basis set with one-center expansions performed inside the MT spheres up to $l_{max}$=6. In the interstitial region the basis functions are expanded in plane waves with the energy up to the cutoff corresponding to 78, 140, and 224 plane waves per *s*, *p* and *d* orbital, respectively. The induced charge densities, the screened potentials and the envelope functions are represented by spherical harmonics up to $l_{max}$ =6 within the MT spheres and by plane waves in the interstitial region with cutoff corresponding to the 16×16×16 fast-Fourier-transform grid in the unit cell of direct space for the bcc structure. The **k**-space integration needed for constructing the induced charge density is performed over the 16×16×16 grid (145 **k** points in the irreducible wedge of the Brillouin zone (BZ)). The improved tetrahedron method was used for the k-point sampling[24].

An earlier first-principle linear response study showed that phonon dispersion in Fe at ambient conditions can be well reproduced by the combined use of the generalized gradient approximations (GGA), spin polarization and ultrasoft pseudopotentials, and the local spin density approximation (LSDA) phonon frequencies are systematically higher than experimental data[17]. In the present work, we use the Perdew-Burke-Ernzerhof (PBE) GGA[25] for the exchange and correlation energy.

B. Electronic contribution

The electronic free energy can be written as

$$F_{el}(V,T) = E_{el}(V,T) - TS_{el}(V,T) \qquad (2)$$

The electronic entropy is given:

$$S_{el}(V,T) = -2k_B \sum_i f_i \ln f_i + (1-f_i)\ln(1-f_i) \qquad (3)$$

Where

$$f_i = \frac{1}{1+\exp\left(\frac{[\varepsilon_i - \mu(T)]}{k_B T}\right)} \qquad (4)$$

is the Fermi-Dirac occupation number. The parameter μ is the chemical potential, $k_B$ is the Boltzmann constant, and $\varepsilon_i$ are the eigenvalues. The sum is over all energy levels, with both the occupied and unoccupied states (up to 1Ry above the static Fermi energy) included. The chemical potential is determined from the particle conservation equation

$$\sum_i f_i = N_{el} \qquad (5)$$

where $N_{el}$ is the total number of the electrons in the systems.

The current method to calculate the electronic free energy $F_{el}$ and electronic entropy $S_{el}$ assumes that the eigenvalues are temperature-independent for given lattice and nuclear positions; only the occupation numbers change with the temperature through the Fermi-Dirac distribution. Wasserman et. al. found that the electronic entropies of hcp Fe calculated from tight-binding total energy calculations based on static eigenvalues agree within 1% with the values from self-consistent high temperature Linear Augmented Plane Wave (LAPW) method over the whole temperature (6000-9000K) and volume (40-90 bohr$^3$/atom) ranges[8], which clearly indicates that the eigenvalue approximation is well justified for transition metals.

C. Vibrational contribution

The linear response method based on density functional theory (DFT) and the density functional perturbation theory (DFPT) has already been successfully applied to calculate the lattice vibrational contribution to the free energy and other thermodynamic properties of many materials[26-29]. In the framework of the linear response LMTO method, the dynamical matrix and the phonon linewidths are determined as a function of wave vector for a set of irreducible **q** points at the 8×8×8 reciprocal lattice grid (29 irreducible **q** points) for the bcc structure. The perturbative approach is employed for calculating the self-consistent change in the potential[28, 29]. Careful tests have been done against **k** and **q** point grids and many other parameters to make sure all the results are well converged. The self-consistent calculation is terminated when the total energy change is less than 10$^{-9}$ Ry in the full potential LMTO, and when the change in the charge density is less than 10$^{-7}$ in the linear response.

The vibrational contribution to the free energy is calculated by combination of the linear response and the quasi-harmonic approximations. Once the phonon dispersion relation and/or phonon density of state is obtained from the linear response lattice dynamics calculations, the phonon internal energy ($u_{ph}$) and phonon free energy ($f_{ph}$) are computed from[32]:

$$u_{ph}(V,T) = \sum_{q,i} \hbar\omega_i(q,V,T)[\tfrac{1}{2} + \frac{1}{e^{\frac{\hbar\omega_i}{k_B T}} - 1}] \quad (6)$$

$$f_{ph}(V,T) = k_B T \sum_{q,i} \ln[2\sinh(\frac{\hbar\omega_i(q,V,T)}{2k_B T})] \quad (7)$$

Within the quasiharmonic approximation for insulators, at a given **q** vector, the frequency ω is solely a function of volume and temperature independent. Although ω does depend on temperature for transition metals because of the electron-phonon coupling, the normal quasiharmonic treatment will not cause any serious problems because the thermal electronic excitation is usually small[26, 27]. In the present paper, we assume that the phonon frequencies ω at a fixed volume are independent of the temperature.

## 3. Lattice dynamics

We calculate lattice dynamical properties of ferromagnetic bcc Fe at 6 different volumes, 65, 75, 75, 79.6, 85 and 90 bohr$^3$/atom. For bcc Fe at volumes significantly beyond the above regimes, some phonon branches will soften and become unstable, consistent with earlier calculatations[13]. In Fig. 1 we show our calculated phonon dispersion curves (solid lines) of ferromagnetic bcc iron at the experimental equilibrium volume (79.6 bohr$^3$/atom) at ambient conditions. For both the transverse (TA) and longitudinal (LA) acoustical branches along several high-symmetry directions in the Brillouin zone, the theoretically determined phonon frequencies agree well with experimental data from recent inelastic neutron scattering measurements at 300K (dots)[18]. There are no phonon anomalies. Dal Corso and Gironcoli performed first principles linear response calculations on magnetic bcc Fe using an ultrasoft pseudopotential; their computed phonon dispersion curves[17] using PBE GGA and LDA approximations are shown as the dashed and dotted lines in Fig. 1. Using linear response and the same PBE GGA approximations for exchange and correlation functional, their ultrasoft pseudopotential and our full potential LMTO calculations are in excellent agreement at ambient conditions. Both also agree well with the experiment. On the other hand, the symmetrical discrepancies between the LDA pseudopotential calculations and the experiment suggests that the LDA approximation not only fails to give the correct ground state structure for Fe[33], but also fails to accurately describe the phonons.

In figure 2 we show the calculated phonon dispersion curves at two compressed volumes: V= 75 bohr$^3$/atom (solid lines) and V=70 bohr$^3$/atom (dotted lines). The calculated phonon frequencies show a significant increase of 6%-15% when the volume decreases from 79.6 to 75.0 bohr$^3$/atom. This is consistent with the inelastic neutron scattering experiment, where all measured phonon frequencies increase 5%-10% for a volume reduction corresponding to 5%[18]. With further volume compression from 75 to 70 bohr$^3$/atom, most phonon frequencies show another 6%-14% increase, except for **q** vectors close to the H point in the BZ boundary, where the increase is much less significant, with only a 1% increase at H point. With increase of pressure, the phonon dispersion shows more complex and anomalous behavior near the H point. At V=75 bohr$^3$/atom, a noticeable dip of the longitudinal mode near [3/4,0,0] appears. The dip becomes more significant at V=70 bohr$^3$/atom, with an additional pronounced dip of both the longitudinal and transverse branches for q vector close to [3/4,3/4,3/4].

Recent developments in high pressure and inelastic neutron scattering techniques by Klotz et. al. have made it possible to measure the lattice vibrational properties under high pressure accurately[34, 35], and their measured phonon dispersion data for bcc Fe at 9.8 GPa[18] are shown as dots in Fig. 2. According to the thermal equation of state (see the following section), the equilibrium volume of bcc Fe at 9.8GPa is close to 75 bohr$^3$/atom, thus we can compare our calculated phonon dispersions at 75 bohr$^3$/atom with the experiment directly. As seen in Fig. 2, the experiment and the calculations agree well at high pressure, often within a few percent over the whole BZ.

Bcc iron transforms under pressure to the hcp phase at around 10~16GPa[36, 37]. Lattice dynamics of bcc Fe under pressure can provide important information in understanding the transition mechanism. According to the well-known Burger mechanism, the bcc to hcp structural transformation can be achieved by two simultaneous distortions: (1) opposite displacements of the bcc (110) planes in the [110] directions, which corresponds to the TA1(N) phonon mode at the bcc BZ boundary[38, 39]; (2) shear deformation in the [001] direction while keeping the volume and the bcc (110) planes unchanged. If the phase transition is dominated by the Burger mechanism, the frequencies of T1(N) phonon mode should show anomalous behavior close to the transition. For the bcc-hcp transition of Ba, first-principles calculations clearly indicate a substantial softening of this T1 N-point phonon mode when approaching the transition pressure (~4GPa)[40]. Such dynamical precursor effects of the lattice instability have also been found in group-IV transition metals such as Ti, Zr and Hf[41-43]. However, the bcc-hcp transition in Fe seems to be quite different. Klotz and Barden measured the phonon dispersion at 0 and 9.8 GPa by inelastic neutron scattering, and demonstrated that such effects are definitely absent[18]. As shown in Fig.1 and 2, instead of softening, compared to the ambient pressure value, the TA1(N) phonon frequency increases ~8% at volume V=75 bohr$^3$/atom (P~10GPa), and ~15% at V=70 bohr$^3$/atom (P~25GPa). For pressures above the bcc to hcp transition pressure, ferromagnetic bcc phase is still dynamically stable, and the dynamic precursor effects according to Burgers mechanism are still absent. The bcc phase eventually becomes dynamically unstable at a much higher pressure (P~180GPa)[13].

The calculated phonon density of states (dos) for ferromagnetic bcc Fe, at five different volumes, V=70, 75, 79.6, 85 and 90 bohr$^3$/atom, are plotted in Fig. 3. With the increase of pressure (decrease of the volume according to the equation of state), the phonon frequencies show a strong increase. The calculated phonon dos at V=79.6 and 75.0 bohr$^3$/atom show a good agreement with the experiment at ambient and high pressure (9.8GPa) [18], respectively.

## 4. Thermal equation of state

The Helmholtz free energies can be evaluated at different volumes and temperatures from the calculated electronic and phonon density of state. The resulting free energies can be further treated in several different ways, as shown in an earlier work on Ta[44].

First, the free energies for each temperature T can be fit to the Vinet equation of state[45-47]:

$$F(V,T) = F_0(T) + \frac{9K_0(T)V_0(T)}{\xi^2}\{1+\{\xi(1-x)-1\}\exp\{\xi(1-x)\}\}]  \quad (8)$$

Where $F_0(T)$ and $V_0(T)$ are the zero pressure equilibrium energy and volume, $x = (V/V_0)^{1/3}$, $K_0(T)$ is the bulk modulus, $\xi = \frac{3}{2}(K_0' - 1)$ and $K_0' = [\frac{\partial K(T)}{\partial P}]_0$. The subscript 0 throughout represents the standard state P=0 GPa.

Pressure can be obtained analytically as:

$$P(V,T) = \{\frac{3K_0(T)(1-x)}{x^2}\}\exp\{\xi(1-x)\} \quad (9)$$

In Fig. 4 we show the calculated thermal equation of state of bcc Fe at temperatures between 250K to 2250K. The calculated thermal equation of state at 250K shows good agreement with diamond-anvil-cell measurements (dots) at the room temperature[48]. While the bcc phase is dynamically stable over the whole pressure and temperature ranges shown here, it is only thermodynamically stable over a small region. Ferromagnetic bcc iron transforms to the fcc phase at T≈1150K, and to the hcp phase at G≈11GPa. There are several reasons to include results beyond this thermodynamically stable field. Bcc phase is metastable in these regions, which might be approached in some shock experiment. The bcc phase becomes stable just before melting at zero pressure, and could become entropically stabilized again at very high temperatures and pressures, although this seems not to happen in pure iron[13]. The computed lattice dynamical and thermal equation of state of bcc phases in these regions may provide crucial information in understanding the phase diagram at extreme conditions.

We show the Vinet equation of state fitting parameters, $F_0(T)$, $V_0(T)$, $K_0(T)$, and $K_0'(T)$ as functions of temperature in Fig.5 and Table 1. Although both LAPW[33] and LMTO calculations underestimate the experimental equilibrium volume[48], GGA shows a dramatic improvement over the LSDA results.

The thermal pressure can be obtained from the pressure difference between isotherms. The thermal pressures as function of volume and temperature are shown in Fig. 6. At low temperature, the thermal pressures are small, and show little volume-dependence. At elevated temperature, the magnitude of thermal pressure increase significantly, and their values first show a decrease with increase of pressure, and then show a strong increase for volumes smaller than 70 bohr$^3$/atom. This is different from bcc Ta, where the volume dependence of the thermal pressure is weak up to 80% compression for a large temperature range (947-9947K)[44]. The different thermal pressure behavior of bcc Fe might be partly due to the pressure dependence of the magnetic moment.

At a given volume, the thermal pressure shows a linear increase with temperature. The pressure change at a given volume is:

$$P(V,T) - P(V,T_0) = \int_{T_0}^{T} \alpha K_T dT \tag{10}$$

where α is the thermal expansion coefficient and $K_T$ is the bulk modulus. For many materials, $αK_T$ is constant in the classical regime. For bcc Ta, for almost all the volumes, the thermal pressure has a slope of ~0.00442GPa/K[44]. However, the slopes for bcc Fe show a strong volume dependence, which might be attributed to its different magnetic moments with volumes[33].

At low temperature, the Helmholtz free energy in the Debye approximation is[19],

$$F = E_{static} + RT[\frac{9}{8}(\frac{\theta_D}{T}) + 3\ln(1 - e^{-\theta_D/T}) - D(\frac{\theta_D}{T})] \qquad (11)$$

Debye function $D(\theta_D/T)$ is

$$D(\frac{\theta_D}{T}) = 3(\frac{T}{\theta_D})^3 \int_0^{\theta_D/T} \frac{z^3 dz}{e^z - 1} \qquad (12)$$

In the classical regime, for temperatures above Debye temperature $\theta_D$, an accurate high-temperature global equation of state can be formed from the T=0K Vinet isotherm plus a volume-dependent thermal free energy $F_{th}$[44]:

$$F_{th} = \sum_{i=1,j=0}^{i=3,j=3} A_{ij} T^i V^j - 3k_B T \ln T \qquad (13)$$

The term TlnT is necessary to give the proper classical behavior at low temperature. A global fit to the calculated thermal Helmholtz free energies can be performed to determine the parameters $A_{ij}$. The above thermal free energy functional gives a good description of various thermal equation of state parameters for bcc Ta at high temperatures[44].

Debye temperatures $\theta_D(T)$ at 0K can be calculated by numerical integration of phonon density of state, and can be obtained according to equation (11) for other temperatures. In Fig. 7 we show the calculated $\theta_D(T)$ as a function of temperature at several different volumes. With increase of pressure (decrease of volume), $\theta_D(T)$ shows a strong increase. At low temperature, $\theta_D(T)$ drops with increase of T. However, with further increase (T > 250K), $\theta_D(T)$ shows almost no temperature dependence. All these are consistent with recent neutron scattering experiment[18]. The calculated $\theta_D(T)$, at both ambient pressure (V=79.6 bohr$^3$/atom) and 9.8 GPa (V=75 bohr$^3$/atom), usually agrees within 10-15K (~2-3%) with the measured data.

The calculated and fitted thermal free energies as function of temperature and volume are compared in Fig. 8. At both low (a) and high (b) temperature regimes, the fit gives good agreement with the calculated data, with rms deviations of ~0.2 mRy. For the high temperature fitting, the residuals (Fig. 8(c)) are small over the whole temperature range for all the volumes studied, with values less than 0.6 mRy. This is different from bcc Ta, where the residuals are much larger due to the electronic topological transition[44].

The various thermal equation of state properties can be derived analytically from the Helmholtz free energy. The thermal expansion coefficient α is:

$$\alpha = \frac{1}{V}(\frac{\partial V}{\partial T})_P = -\frac{1}{V}(\frac{\partial^2 F}{\partial T \partial V})/(\frac{\partial^2 F}{\partial V^2})_T \qquad (14)$$

At low temperatures, the Debye expression for α is[19]:

$$\alpha = \frac{3R\gamma_D}{K_T V}[4D(\frac{\theta}{T}) - \frac{3(\theta/T)}{e^{\theta/T}-1}] \qquad (15)$$

The thermal expansion coefficient of bcc Fe agrees well with experiment at low temperatures [Fig. 9(a)]. At high temperature, α shows a linear increase under several different pressures [Fig. 9(b)], as predicted by the quasiharmonic approximation at the high T limit[19]. Thermal expansivity is a very sensitive parameter, and the discrepancy between the calculated α at ambient pressure and high temperature with the experiment[49,50] might be attributed to several factors: the errors in the first-principles calculations, anharmonic effects, and most likely magnetic fluctuations. It should be noted that the calculated α shows good agreement with experiment even at high temperature when applying similar theoretical approaches to nonmagnetic bcc vanadium.

The calculated thermal expansion coefficient shows a rapid drop with increasing pressure [Fig. 9(c)]. The relationship between α and pressure are characterized by the Anderson-Grüneisen parameter $\delta_T$[19]:

$$\delta_T = (\frac{\partial \ln \alpha}{\partial \ln V})_T = -\frac{1}{\alpha K_T}(\frac{\partial K_T}{\partial T})_P \qquad (16)$$

The calculated $\delta_T$ of bcc Fe shows quite complex behavior as a function of pressure and temperature [Fig. 10]. At a given pressure, $\delta_T$ first decreases with temperature, and then shows a slight increase, similar to the behavior in bcc Ta[44]. At all temperatures, $\delta_T$ shows a strong decrease with pressure. For many materials, the parameter $\delta_T$ can be fitted to a form as a function of volume[51]:

$$\delta_T = \delta_T(\eta=1) \times \eta^{\kappa} \qquad (17)$$

where $\eta = V/V_0(T_0)$. For bcc Ta, the average $\delta_T$ shows $\delta_T(\eta) = 4.56 \times \eta^{1.29}$ for temperature 0-6000K[44], and $\delta_T(\eta) = 4.56 \times \eta^{1.29}$ has been reported for MgO at 1000K[51]. However, bcc Fe shows different behavior. Although the parameter $\delta_T$ shows a strong decrease with compression at all temperatures, it does not drop that rapidly as power order when the pressure is high [Fig. 10(c)]. Similar behavior has also been reported for fcc and hcp Fe[8].

The Grüneisen ratio γ is an important thermodynamic parameter used to quantify the relationship between the thermal and elastic properties of a solid, particularly for understanding shock dynamics[19].

$$\gamma = V(\frac{\partial P}{\partial U})_V = \frac{\alpha K_T V}{C_V} = V\frac{\partial^2 F}{\partial V \partial T} \Big/ (\frac{\partial U}{\partial T})_V \qquad (18)$$

where U is the internal energy. As shown in Fig. 11, at a given pressure, γ of bcc Fe first shows an increase with increasing temperature, and then a rapid decrease when T > ~ 1000K. On the other hand, the variation of γ with pressure is moderate. This is significantly different from bcc Ta, where the temperature dependence is moderate, but the pressure dependence is not.

The calculated Grüneisen ratio at 500K shows good agreement with experiment using the adiabatic decompression method[52]. The volume dependence of the Grüneisen ratio is given by the parameter $q$:

$$q = \frac{\partial \ln \gamma}{\partial \ln V} \qquad (19)$$

The parameter $q$ is often assumed to be a constant, for example, 0.6 for bcc Fe[52], and 0.7-1.62 for hcp Fe depending on the pressure range and measuring methods[53]. However, our calculations show that $q$ is both temperature and pressure dependent [Fig. 12]. The parameter $q$ decreases significantly with pressure, but its temperature behavior is quite complex. At ambient pressure, $q$ shows a slight increase with temperature. However, at high pressure, $q$ first shows a strong decrease with increasing temperature, and then a slight increase when T > ~ 1500K. Similar complex behavior the parameter $q$ has also been reported for bcc Ta[44].

At low temperatures, the Debye heat capacity at constant volume $C_V$ is:

$$C_V = 3R[4D(\frac{\theta}{T}) - \frac{3(\theta/T)}{e^{\theta/T} - 1}] \qquad (20)$$

and is in good agreement with experiment [Fig. 13(a)]. At high temperatures [Fig. 13 (b) and (c)], $C_V$ is pressure independent at a given temperature. When the temperature is less than ~1200K, $C_V$ only shows slight increase with temperature. At higher temperature, the increase of $C_V$ becomes more noticeable. This increase comes mainly from the electronic excitation contribution.

# 5. Conclusions

In summary, we have performed detailed first principles linear response calculations to study the lattice dynamics and thermal equation of state properties of ferromagnetic bcc Fe. The calculated phonon dispersion and phonon density of state, both at ambient and high pressures, agree well with inelastic neutron scattering experiment. No dynamic precursor effects of lattice instability exist for the bcc-hcp phase transition. The calculated free energies have been treated by three different forms: Vinet equation of state, simple linear thermal pressure and a global fit to the thermal Helmholtz free energy. The calculated thermal equation of state agrees well with experiment. The thermal expansion coefficient agrees well with the experiment at low temperature. The difference of high temperature might be attributed to the influence of magnetic fluctuations. The calculated Grüneisen ratio and heat capacity $C_V$ show little pressure dependence. Thermal electronic excitations contribute significantly to the temperature dependence of $C_V$ at high temperature.

**ACKNOWLEDGEMENTS**

Much thanks to S. Y. Savrasov for kind agreement to use his LMTO codes and many helpful discussions. This work was supported by DOE ASCI/ASAP subcontract B341492 to Caltech DOE w-7405-ENG-48. Computations were performed on the Opteron Cluster at the Geophysical Laboratory, supported by the above DOE grant.

**FIGURES**

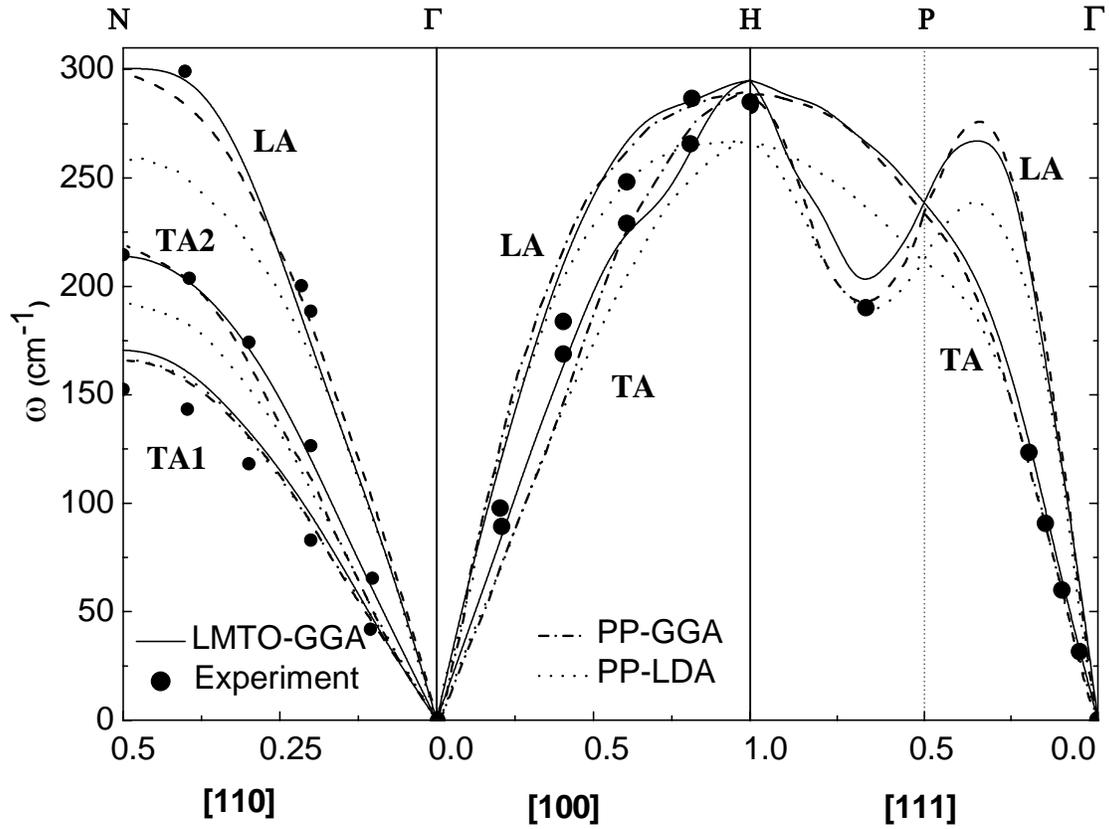

Figure 1. Calculated phonon dispersion curves (solid lines) for ferromagnetic bcc Fe at 79.6 bohr$^3$/atom, in comparison to neutron inelastic scattering measurements (dots, Ref.18 ) and first principles linear response calculations using ultrasoft pseudopotential with GGA (dashed lines) and LDA (dot lines) approximations computed at experimental lattice constant (Ref. 17).

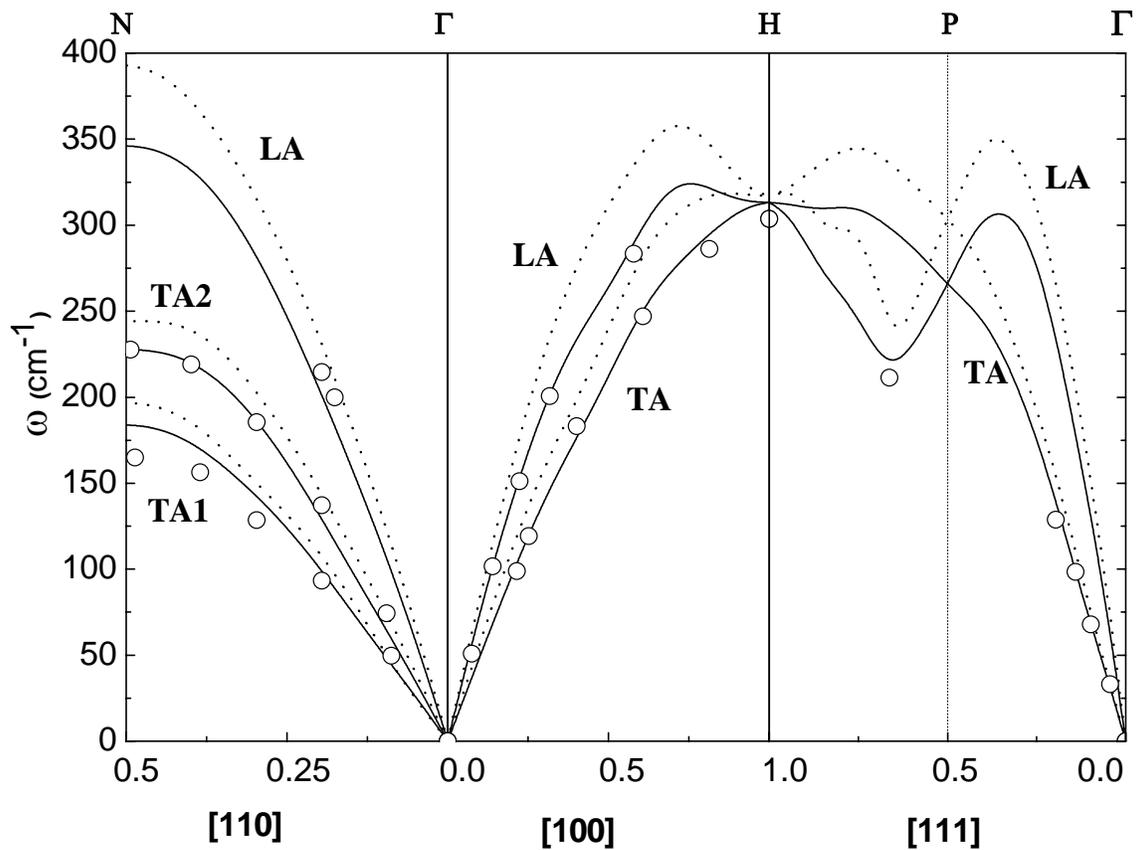

Figure 2. Calculated phonon dispersion curves for ferromagnetic bcc Fe at two compressed volumes, V=75 bohr$^3$/atom (solid lines) and V=70 bohr$^3$/atom (dotted lines), in comparison to the neutron inelastic scattering data at 9.8 GPa (dots, Ref. 18).

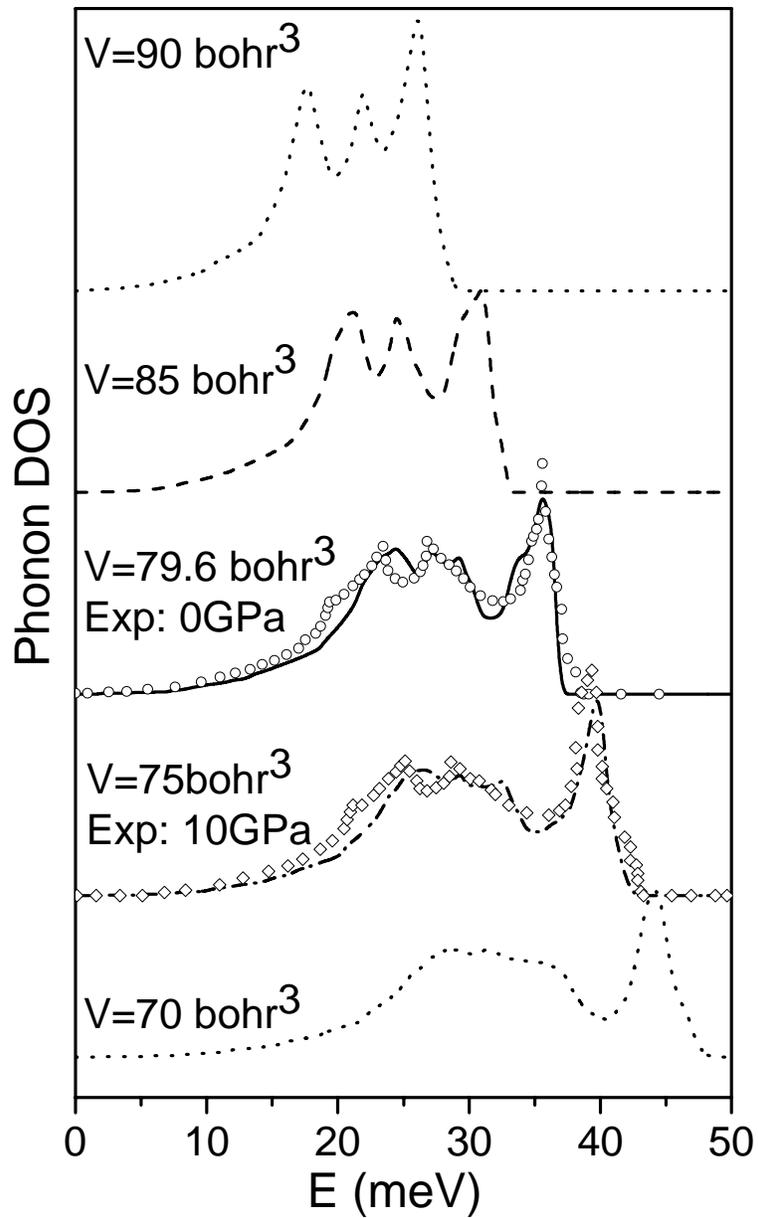

Figure 3. Calculated phonon density of states (dos, lines) of ferromagnetic bcc Fe at five different volumes, 70, 75, 79.6, 85 and 90 bohr$^3$/atom, in comparison to the neutron inelastic scattering data at 0 and 9.8 GPa (dots, Ref. 18).

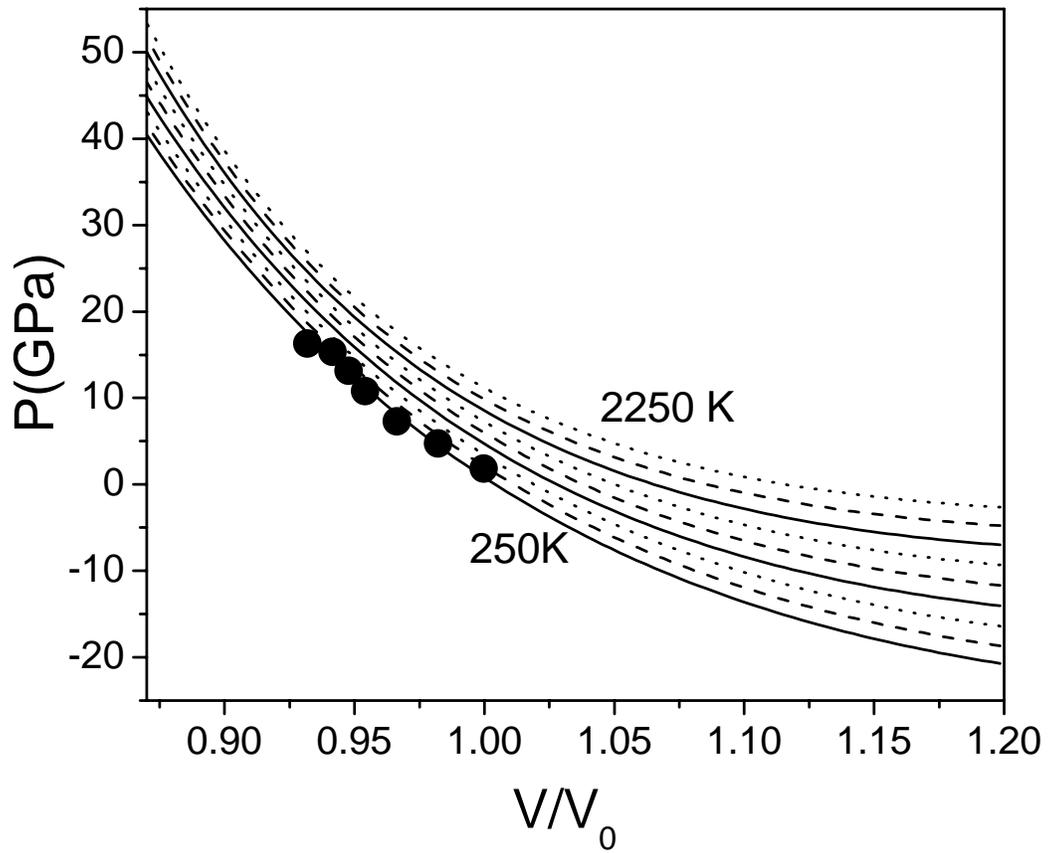

Figure 4. Calculated thermal equation of state (lines) at temperatures between 250K and 2250K. Experimental equation of state, measured in a diamond anvil cell at the room temperature, is shown as dots (Ref. 48).

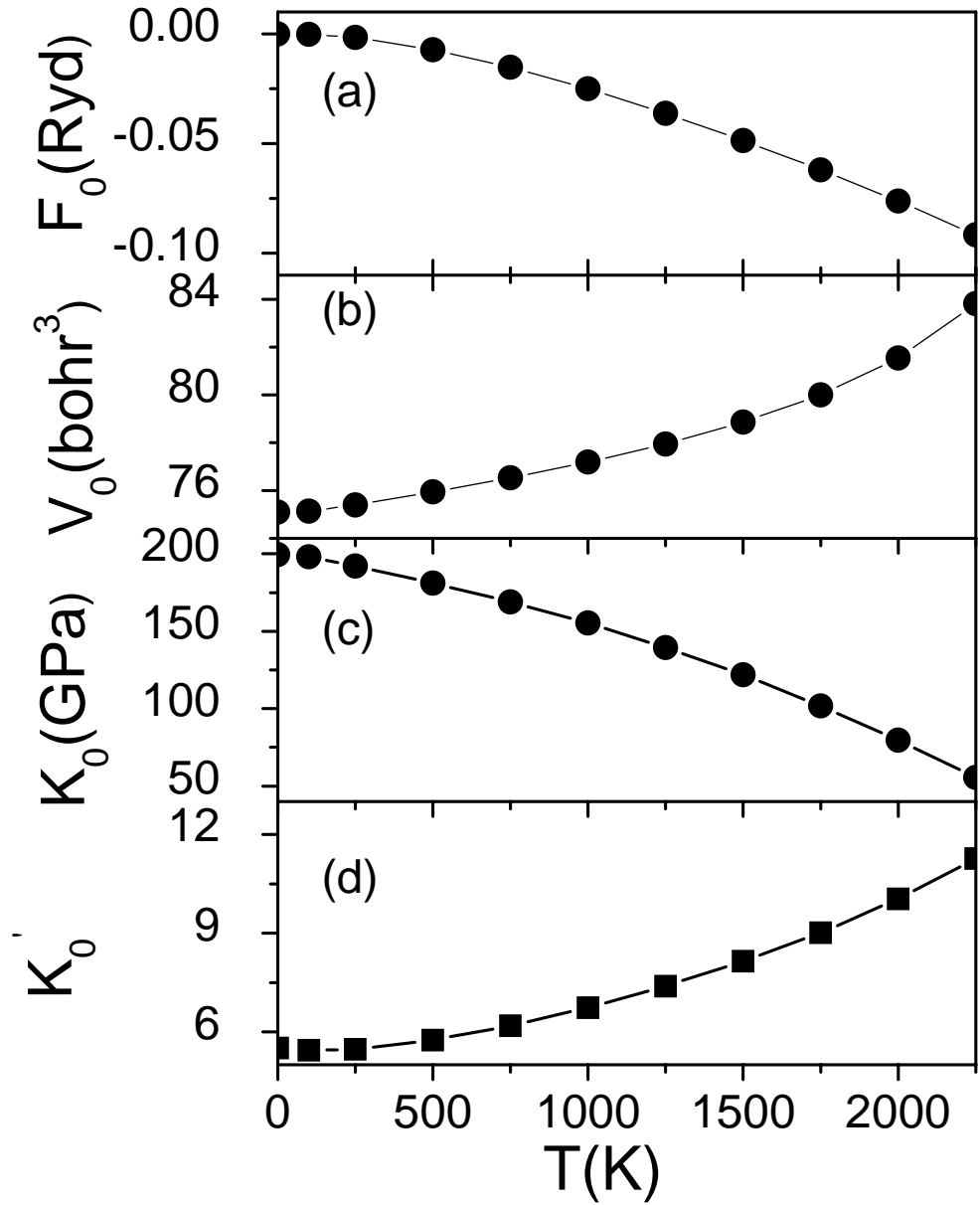

Figure 5. Fitted Vinet equation of state parameters, $F_0(T)$ (a), $V_0(T)$ (b), $K_0(T)$ (c), and $K_0'(T)$ (d) as function of temperature.

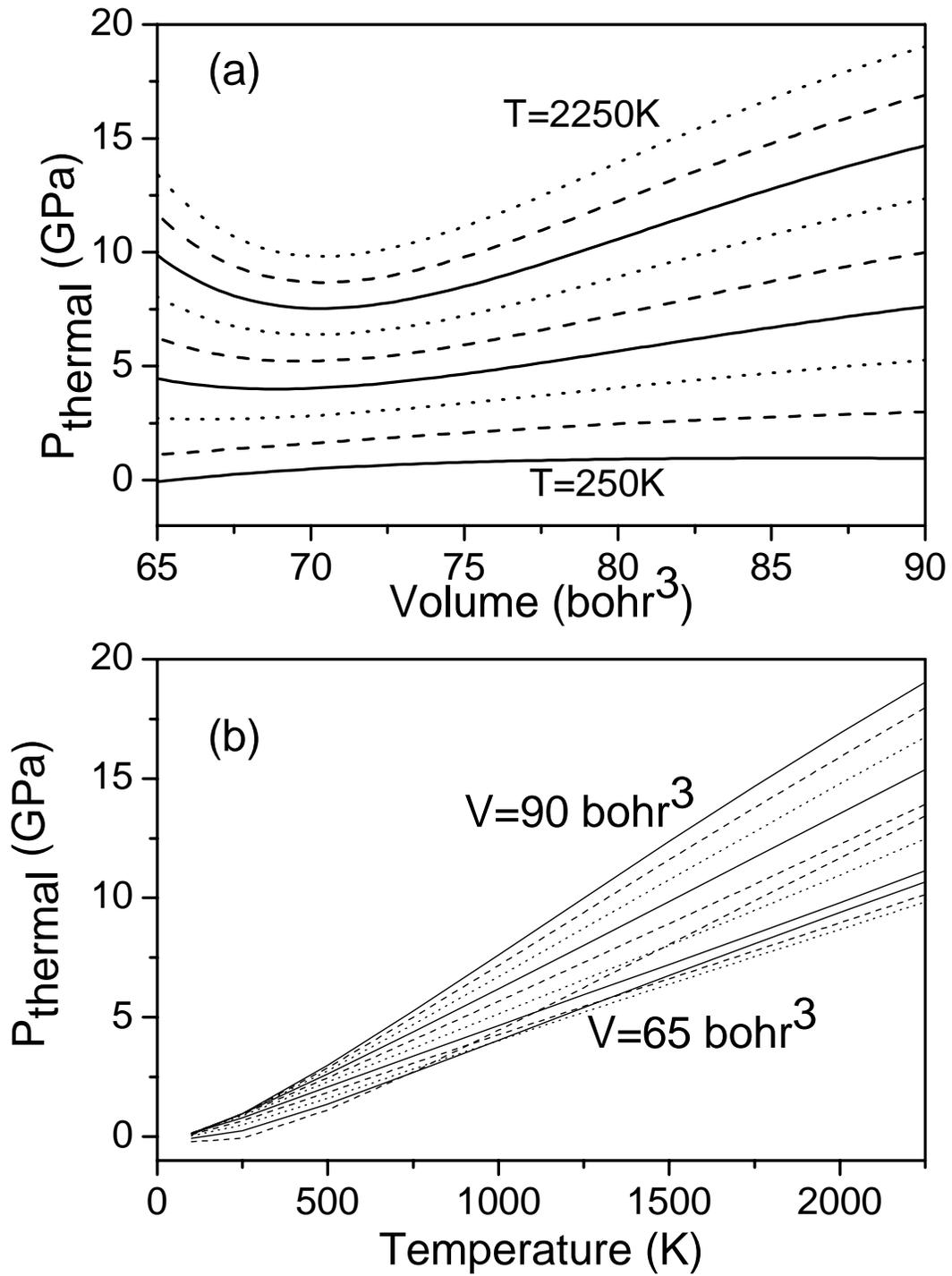

Figure 6. Thermal pressure of bcc Fe as function of volume (a) and temperature (b).

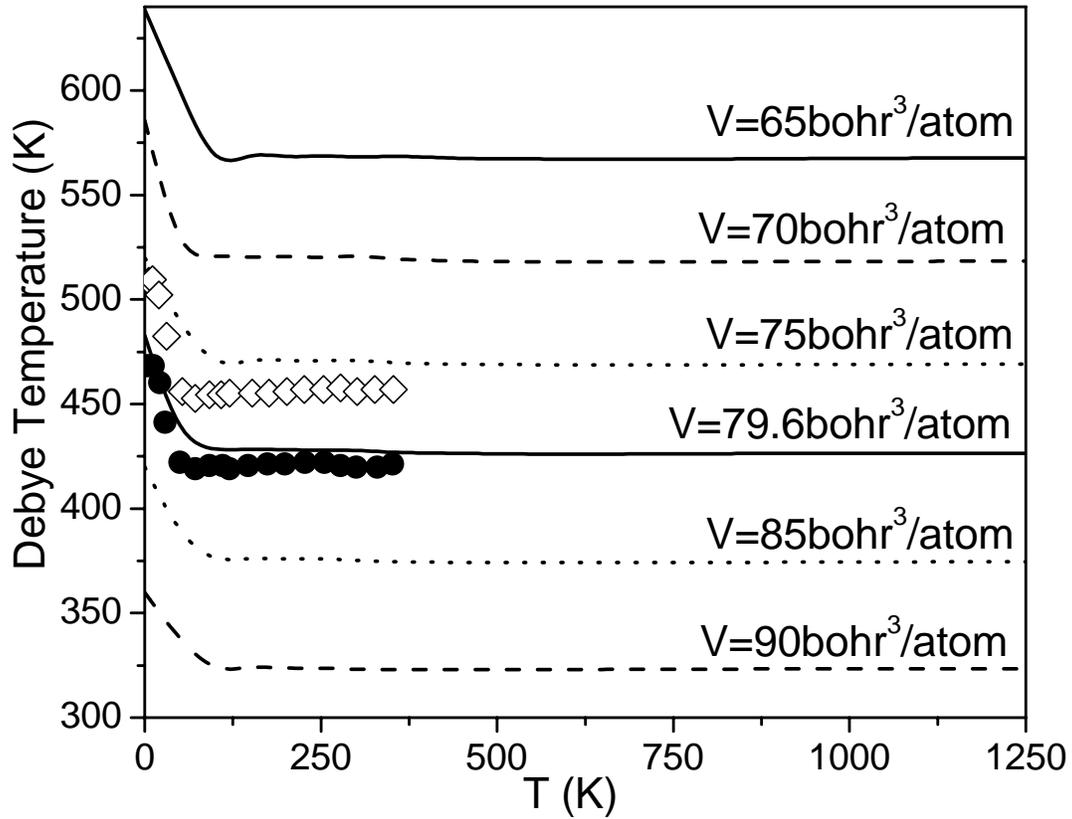

Figure 7. Calculated Debye temperature $\Theta_D(T)$ as a function of temperature at several volumes. $\Theta_D(T)$ shows a rapid increase with pressure (decrease of volume). With increase of temperature, $\Theta_D(T)$ drops rapidly at low temperature, but shows little temperature dependence when T>250K. Neutron scattering experimental $\Theta_D(T)$ at 0 and 9.8 GPa are shown as filled circles and open diamonds, respectively (from Ref. 18).

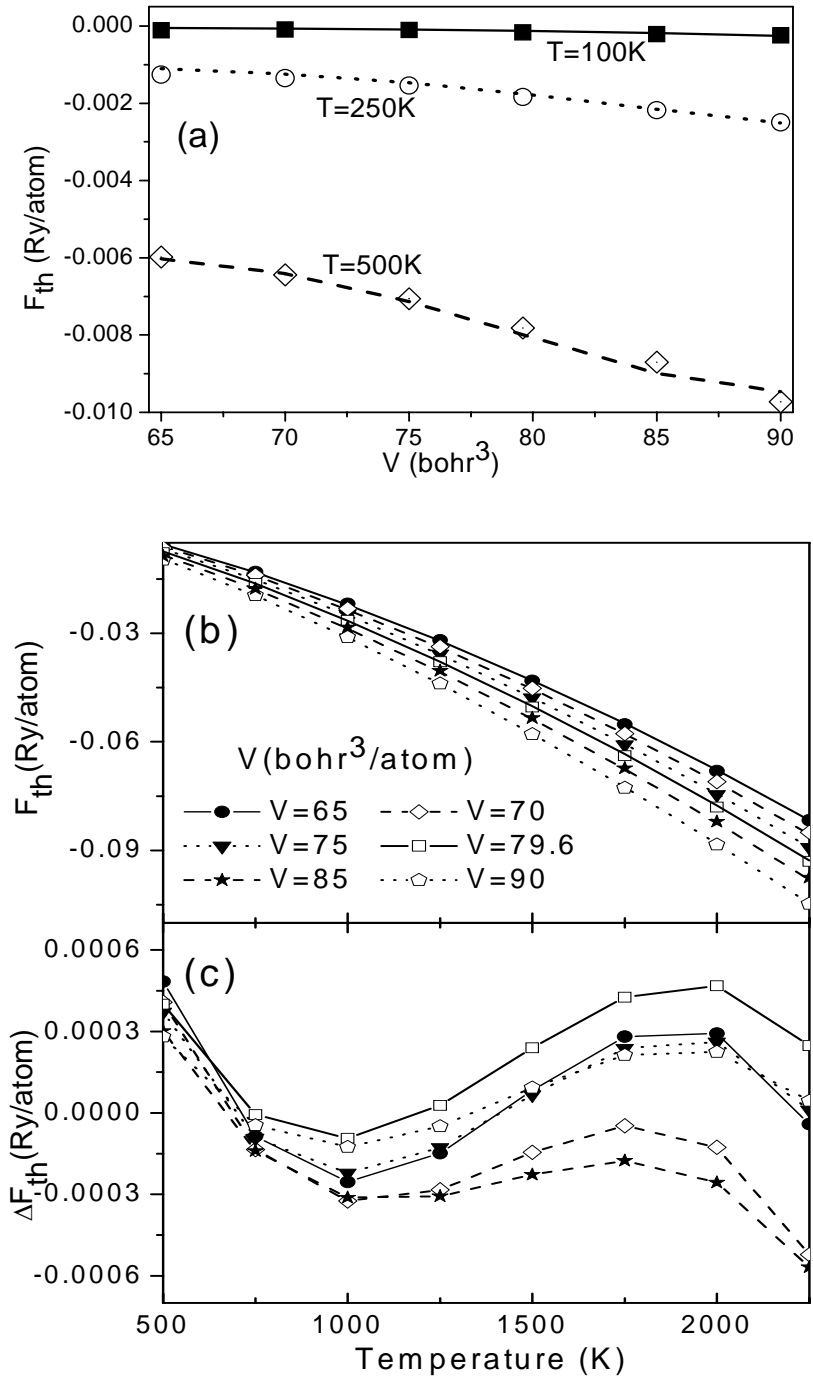

Figure 8 Global fit of thermal free energies to the calculated first principles data at several different volumes and temperatures. At both low (a) and high (b) temperature regimes, the fitted (lines) thermal energies agrees well with the computed data (symbols); and the residuals at high temperatures(c) are very small.

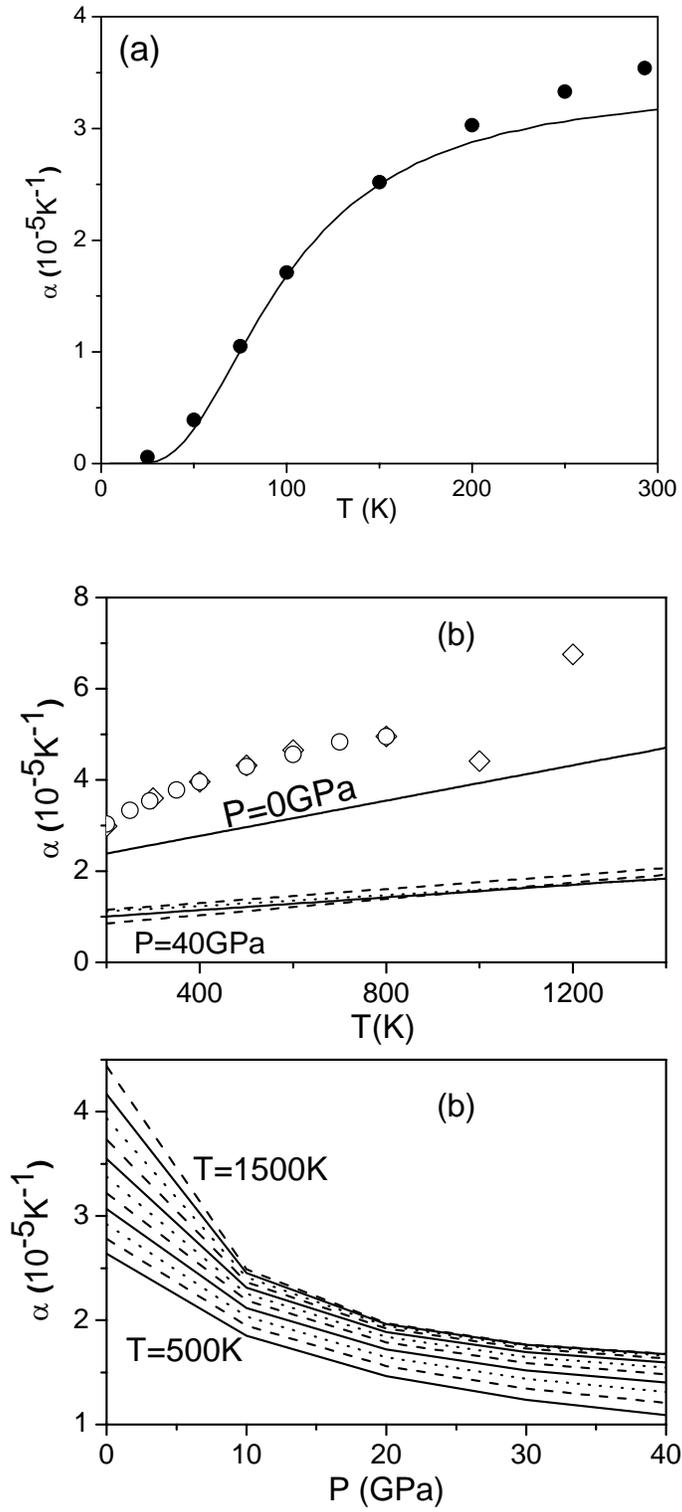

Figure 9 The thermal expansion coefficient of bcc Fe as function of temperature (a and b) and pressure(c). Experimental data (dots) are from Ref. 49 and 50.

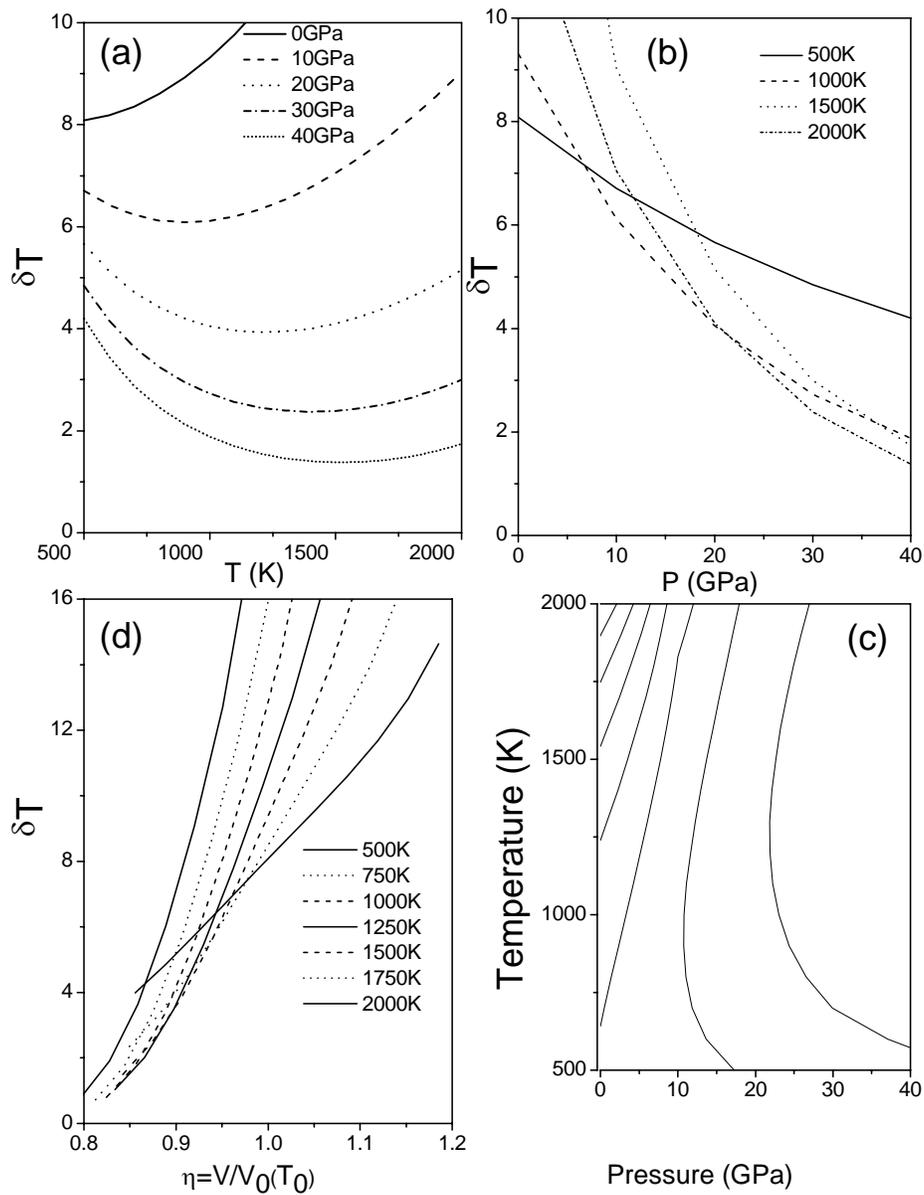

Figure 10 The Anderson-Grüneisen parameter $\delta_T$ as a function of (a) T, (b) P (d) $\eta= V/V_0$ (T=0) (c) countours of $\delta_T$.

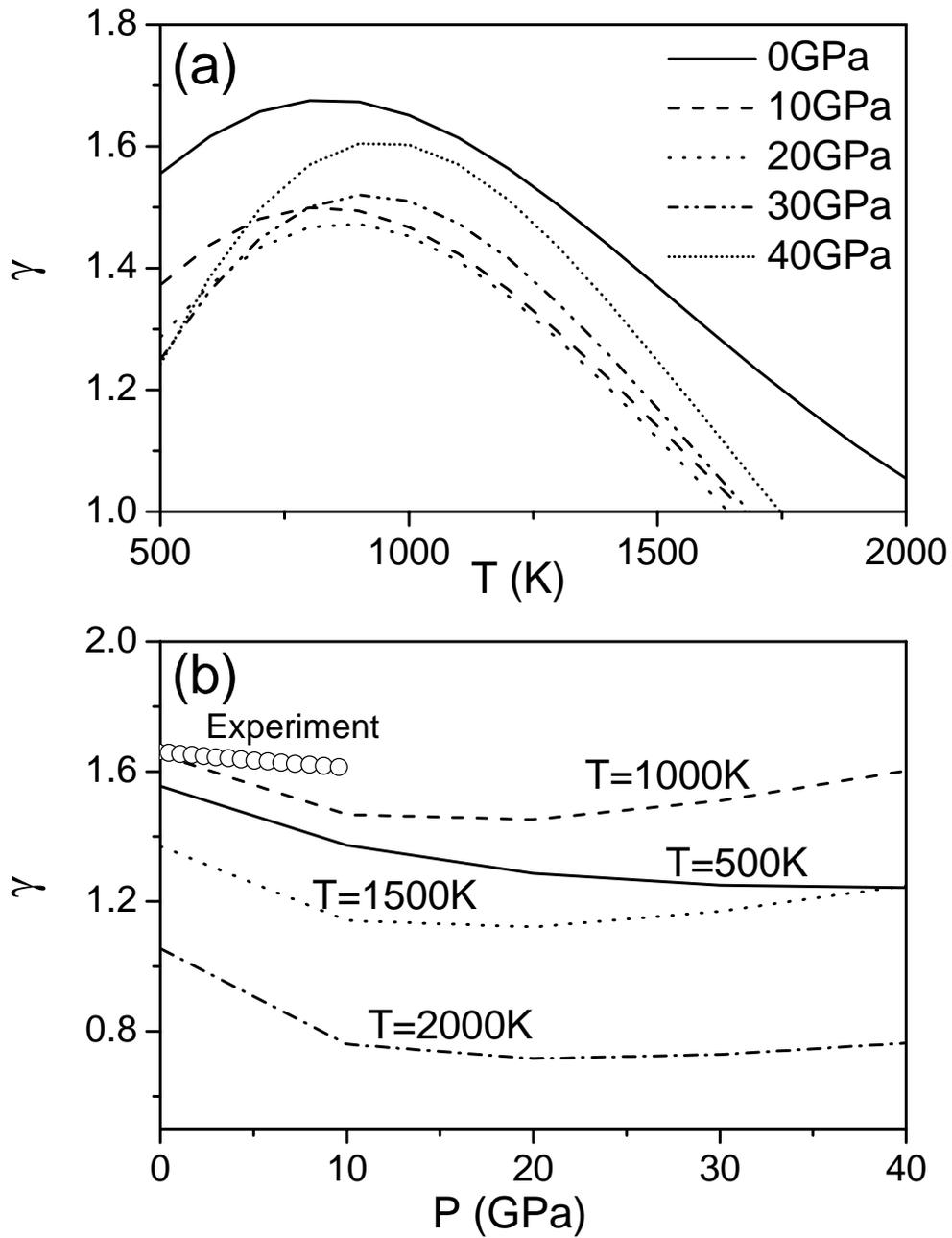

Figure 11 The Grüneisen ratio γ as a function of (a) temperature and (b) pressure. Experimental data (dots) are from Ref. 53.

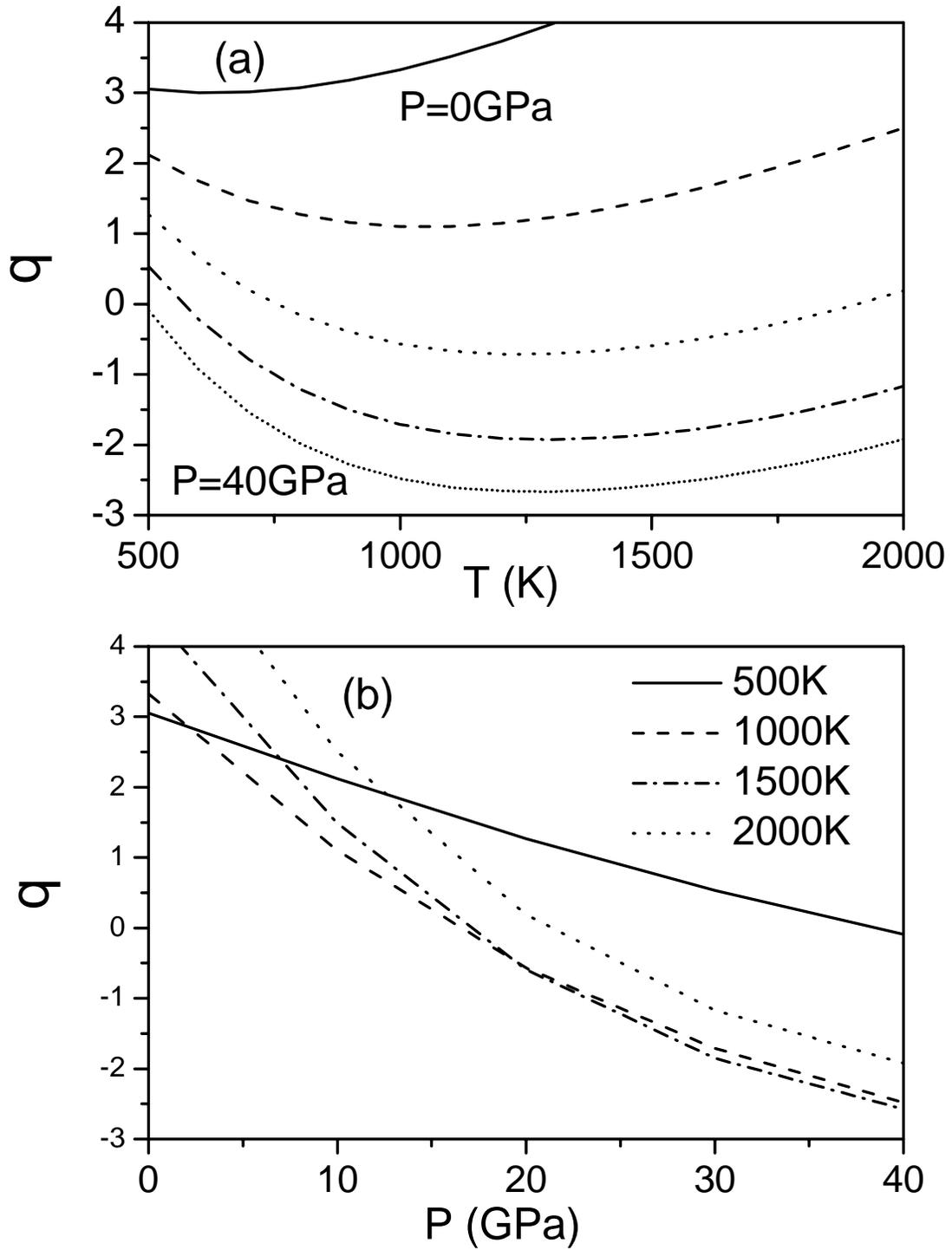

Figure 12 Varation of the parameter *q* as a function of (a) temperature and (b) pressure.

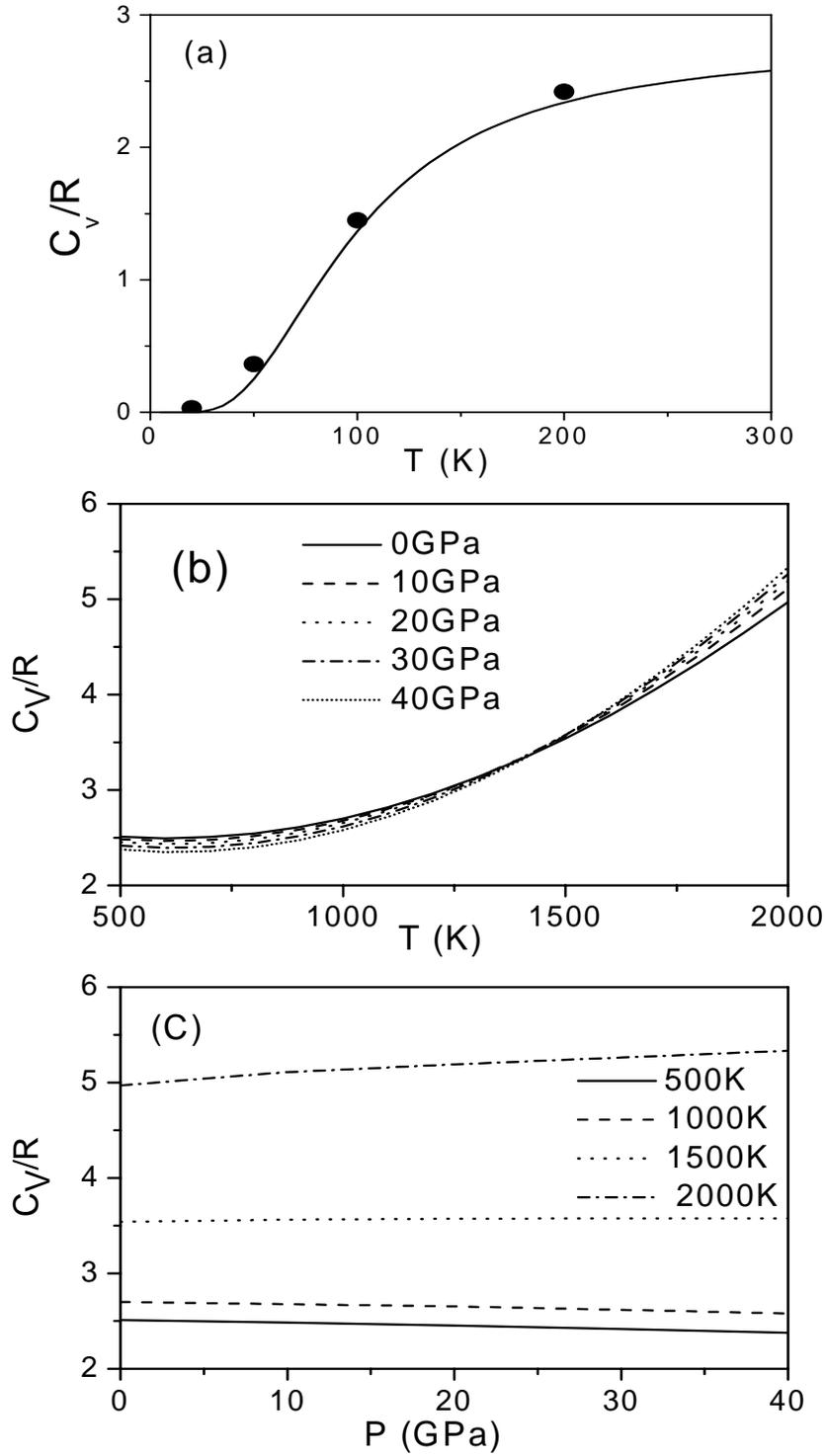

Figure 13  Heat capacity at constant volume. Calculated $C_V$ agrees with experiment (dots, Ref 49) at low temperature (a). At high temperature, $C_v$ increases significantly due electronic thermal excitation(b), but shows little pressure dependence (c).

Table 1  Equation-of-state parameters determined by fitting a Vinet equation to the computed binding energy curves. LAPW results are from Ref. 33, and experimental data are from Ref. 48.

|           | $V_0$ (bhor$^3$) | $K_0$ (GPa) | $K_0'$ |
|-----------|------------------|-------------|--------|
| LMTO-GGA  | 75.36            | 178         | 4.7    |
| LAPW-GGA  | 76.84            | 189         | 4.9    |
| LAPW-LSDA | 70.73            | 245         | 4.6    |
| Expt.     | 79.51            | 172         | 5.0    |